# Dynamic interfaces in an organic thin film

Chenggang Tao[1,2], Qiang Liu [1,2], Blake C. Riddick[1,2], William G. Cullen[1,2], Janice Reutt-Robey[1,3], John D. Weeks[1,3,4] and Ellen D. Williams[1,2,3,4]

[1] *Materials Research Science and Engineering Center, University of Maryland, College Park, Maryland 20742*

[2] *Department of Physics, University of Maryland, College Park, Maryland 20742*

[3] *Department of Chemistry and Biochemistry, University of Maryland, College Park, Maryland 20742*

[4] *Institute for Physical Science and Technology, University of Maryland, College Park, Maryland 20742*

**Abstract**

Low-dimensional boundaries between phases and domains in organic thin films are important in charge transport and recombination. Here, fluctuations of interfacial boundaries in an organic thin film, acridine-9-carboxylic acid (ACA) on Ag(111), have been visualized in real time, and measured quantitatively, using Scanning Tunneling Microscopy. The boundaries fluctuate via molecular exchange with exchange time constants of 10-30 ms at room temperature, yielding length mode fluctuations that should yield characteristic $f^{-1/2}$ signatures for frequencies less than ~100 Hz. Although ACA has highly anisotropic intermolecular interactions, it forms islands that are compact in shape with crystallographically distinct boundaries that have essentially identical thermodynamic and kinetic properties . The physical basis of the modified symmetry is shown to arise from significantly different substrate interactions induced by alternating orientations of successive molecules in the condensed phase. Incorporating this additional set of interactions in a lattice gas model leads to effective multi-component behavior, as in the Blume-Emery-Griffiths (BEG) model, and can straightforwardly reproduce the experimentally observed isotropic behavior. The general multi-component description allows the domain shapes and boundary fluctuations to be tuned from isotropic to highly anisotropic in terms of the balance between intermolecular interactions and molecule-substrate interactions.

Key words:  Organic thin film, fluctuations, STM, molecular interactions, diffusion kinetics, phase coexistence

**Introduction**

Organic materials that allow electron and hole transport and recombination have extensive materials and device applications (1-3). Understanding the relationships between the transport characteristics of organic thin films (OTF) and how molecules are structurally arranged within the film is a fundamental and challenging issue. For OTFs, complex molecular symmetries and interactions (4-6) often cause the formation of multiple density-dependent phases, with pattern size on the nanometer or micron scale, coexisting at room temperature (7-9). The boundaries of such local ordered regions (often referred to as domain boundaries or island edges) may dramatically affect the stability, transport and charge recombination properties of the OTFs (10-13). We report observations of the dynamic behavior of such low-dimensional interfaces, obtained using time-dependent scanning tunneling microscopy (STM) to image the boundaries between the ordered and disordered phases of acridine-9-carboxylic acid (ACA) on Ag(111) at room temperature.

The ACA molecule, shown in Fig. 1a, has a nominally rectangular structure, with the potential for strong hydrogen-bond (H-bond) intermolecular interactions along the short axis, and weaker quadrupolar interactions along the long axis. Our previous STM studies of ACA adsorption have shown that at low to moderate ACA coverage (< 0.3 ML), no ordered ACA structures form on the terraces of the substrate (14-16). As ACA coverage is increased beyond 0.3 ML, ordered areas (islands) begin to form, covering the substrate partially. The area around the islands shows no ordered overlayer at room temperature, but noise in the tunneling current suggests the presence of mobile species. This is confirmed by measurements at low temperature, where the motion is quenched (14). Thus above 0.3 ML, ACA exists in two-phase equilibrium between a dense ordered phase and a disordered phase. High resolution imaging (15, 16), see Fig. 1b, shows that alternating ACA molecules are inequivalent, e.g. there are two ACA molecules without mirror symmetry per unit cell. XPS core level measurements (15, 16) reveal a hydrogen-bonding interaction between ACA molecules, with the ring nitrogen acting as the H-bond acceptor and the carboxyl proton acting as the H-bond donor, indicating the molecular arrangement (15, 16) illustrated in Figs. 1c-d. At moderate ACA surface densities, ACA molecules arrange head-to-tail into chains connected by H-bonds between the carboxyl group on one side of the ACA molecule, and the ring nitrogen on the other side. These chains run along the $[1\bar{1}0]$ direction of the substrate, as shown in Fig. 1b. The asymmetry of the ACA ordered structure provides two distinct types of edge boundaries (parallel = S and perpendicular = CE to the hydrogen-bonded rows), which are the subject of this report.

In the simplest lattice model of the ordered overlayer, the anisotropic ACA interactions alone would determine the lateral order, with the substrate defining only the lattice constant and crystallographic directions. In this picture, the CE boundaries would have larger edge energies due to the broken H-bonds (magnitude typically ~0.3 eV per bond) and easy excitation of edge roughness due to the weak quadrupolar interactions (magnitude typically 4 to 6 times smaller than the H-bonds) along the edge (17, 18). The energy/roughness behavior would be the opposite for the S boundaries. However, as will be shown here, quantitative analysis of the edge behavior reveals essentially identical behavior of the two types of edge boundaries, ruling out such a simple lattice model. By modifying the lattice-gas approach (19), we will show that the results reveal a strong non-uniformity of the molecule-substrate interaction for successive molecules in an H-bonded pair, a result with significant implications for interpretation of the structures of molecular overlayers in general.

**Experimental Results and Analysis**

Images of ordered regions formed at an average ACA coverage of ~ 0.6 ML are shown in Fig. 2. The isolated ordered ACA islands were formed on wide (111) terraces and do not contact silver steps.. Such isolated islands are typically compact, with an aspect ratio of the island width along the $[1\bar{1}0]$ direction of the substrate to the width along the $[11\bar{2}]$ direction near to one. The aspect ratio $r_S/r_{CE}$ of the island, shown in Figure 2, is directly related to the edge formation free-energies $\beta_{CE}$ and $\beta_S$ for the two types of steps shown in Fig. 1, as $r_S/r_{CE} = \beta_S/\beta_{CE}$ (20-23). While more detailed expressions are available (20, 24, 25), at temperatures well below the critical temperature the island shape is reasonably described by its T = 0 value. In the simple lattice-model discussed above, the T = 0 edge energy is determined by the interaction energies $\varepsilon_{S,CE}$ perpendicular to the boundary edge and the molecular lengths $a_{S,CE}$ along the edge as $\beta_{S,CE} \approx \varepsilon_{S,CE}/a_{S,CE}$. Thus the observation of $r_S/r_{CE} \sim 1$ yields $\varepsilon_{CE}/\varepsilon_S \sim \sqrt{3}$, in distinct contrast to the expected value of 4-6 for the ratio of the H-bonding to quadrupolar interaction strength.

The edges of ACA ordered islands, *i.e.* the boundaries between the ordered phase and the disordered phase, as shown in Fig. 3a , appear frizzy (26), clearly indicating thermal fluctuations at room temperature. The boundary shown is *ca.* perpendicular to the $[1\bar{1}0]$ direction of the substrate. Based on the molecular arrangement in the ordered phase (see Fig. 1), this boundary is formed by ends of the ACA chains in the ordered phase, corresponding to the CE boundary of Fig. 1. The inset shows a boundary roughly perpendicular to the $[11\bar{2}]$ direction, corresponding to the S boundary of Fig. 1. For temporal imaging, the STM scan direction is chosen perpendicular to the boundary orientation, as indicated by white arrows. Examples of the temporal pseudo-images formed as the STM tip repeatedly scans over a point at the boundary are shown in Fig. 3b. The vertical axis is time *t* and the horizontal axis is the position of the point *x(t)*. The distribution of displacements has a Gaussian distribution, as required for fluctuations in thermal equilibrium. There is no deviation from the Gaussian shape for scans measured in either direction across the boundary, providing strong confirmation that under the chosen tunneling conditions the STM scans are not perturbing the thermal distribution at the boundaries. The mean squared boundary width $w^2 = \langle (x - \bar{x})^2 \rangle$ is obtained by fitting the Gaussian, yielding width $w = 2.04 \pm 0.04$ nm for CE boundaries and $2.01 \pm 0.08$ nm for S boundaries.

The free energies and time constants governing the behavior of the boundaries can be evaluated from the correlation functions of the boundary fluctuations (27-29). Given the experimentally measured boundary displacements, *x(t)*, it is straightforward to determine the time correlation function *G(t)* and autocorrelation function *C(t)* respectively (28):

$$G(t) = \langle (x(t+t_0) - x(t_0))^2 \rangle, \tag{1}$$

$$C(t) = \langle (x(t+t_0) - \bar{x})(x(t_0) - \bar{x}) \rangle, \tag{2}$$

Each individual *x(t)* data set is used to calculated individual correlation functions. Fig. 4 shows typical results for *G(t)* and *C(t)*, each the average of the individual correlation functions for more than 10 *x(t)* sets.

The continuum step model is used to relate the correlation functions to the physical properties of boundary (27-29). As shown in Fig. 4, the measured time correlation function *G(t)* is well fit using:

$$G_0 = \left(\frac{2\Gamma(1-1/n)}{\pi}\right)\left(\frac{k_B T}{\tilde{\beta}}\right)^{\frac{n-1}{n}} \Gamma_a^{1/n}, \tag{3}$$

where $\Gamma_a$ is the mobility of the boundary, $\Gamma(1-1/n)$ the gamma function and $\tilde{\beta}$ the step stiffness. The fit to the time correlation function, shown as the solid red curve in Fig. 4, yields the exponent $1/n = 0.52 \pm 0.07$. To evaluate the variance of the values, we average 3 independent measurements for the CE boundaries, and 4 independent measurements for the S boundaries. The average value $1/n = 0.52 \pm 0.09$ for the CE boundary and $0.51 \pm 0.05$ for the S boundary. These values show that the boundary fluctuations are dominated by uncorrelated events for both boundaries ($n = 2$) (27, 30, 31), as opposed to correlated events such as diffusion along the boundaries (24, 26, 29, 32-34), which would yield $n = 4$. The fits also yield the magnitudes of the time correlation functions, which are $G_0 = 3.76$ nm$^2$ for CE boundaries and $3.70$ nm$^2$ for S boundaries. Using Eq. (3), we can find the ratios of the physical constants governing the edge fluctuations: $k_B T \Gamma_a / \tilde{\beta} \approx 11.1$ nm$^4$s$^{-1}$ for CE boundaries and $10.8$ nm$^4$s$^{-1}$ for S boundaries. Since the mobility of a step edge is generally lower when the stiffness is larger, it is surprising to find these ratios similar for the two types of step edges. The significance of these values will be discussed below.

Given the result that the boundary fluctuations are predominantly due to molecular exchange ($n = 2$), we can further analyze the autocorrelation function as shown in Fig. 4. For $n = 2$, the autocorrelation functions have the time dependence:

$$C(t) = C(0)\left[e^{-t/\tau_c} - \Gamma(1/2, t/\tau_c)(t/\tau_c)^{1/2}\right], \tag{4}$$

$$\tau_c = \left(\frac{L}{2\pi}\right)^2 \frac{k_B T}{\Gamma_a \tilde{\beta}}, \tag{5}$$

where $\tau_c$ is the correlation time, $\Gamma(1/2, t/\tau_c)$ the incomplete gamma function, $L$ the correlation length (or effective system size), and $\Gamma_a$ the mobility of the boundary. The fitting curve, shown as the solid red curve in Fig. 4 yields the values for $C(0)$ and $\tau_c$. The average value of $C(0)$ is 2.74 nm$^2$ for CE boundaries and 2.42 nm$^2$ for S boundaries. The average correlation times are found to be $\tau_c = 3.59$ s for CE boundaries and 2.73 s for S boundaries, substantially smaller than the measurement time of 102.4 s. The edge length of the island boundaries, typically from 50 nm to 70 nm, was used to estimate the range of values for the system size, $L$, of Eq. (5). The lower limit of $L$ (= edge length) is obtained if we assume that the island edge fluctuations are not constrained by the corners of the island. If we assume that the corners completely pin the fluctuations, the we obtain the upper limit of $L$ (= 2 x edge length) (35). Using the measured time constants and $L =$ 60 nm (120 nm) in Eq. (5) gives the values $k_B T / \Gamma_a \tilde{\beta} = 0.039$ nm$^{-2}$s (0.16 nm$^{-2}$s) for CE boundaries and 0.030 nm$^{-2}$s (0.12 nm$^{-2}$s) for S boundaries. Combining these values with the values of $k_B T \Gamma_a / \tilde{\beta}$ obtained from the values of $G_0$ above we obtain, for CE boundaries, edge stiffness $\tilde{\beta} = 39.1$ meV/nm (78.2 meV/nm) and edge mobility $\Gamma_a =$ 16.8 nm$^3$/s (33.6 nm$^3$/s); and for S boundaries, $\tilde{\beta} = 45.5$ meV/nm (91.0 meV/nm) and $\Gamma_a$ = 19.0 nm$^3$/s (38.0 nm$^3$/s). Within the experimental uncertainties shown in Table 1, the values for the two different boundary types are the same: $\tilde{\beta}_s / \tilde{\beta}_{CE} = 1.2 \pm 0.3$ and $\Gamma_{a,S} / \Gamma_{a,CE} = 1.1 \pm 0.1$.

The measured stiffness values can be used in a simple lattice approximation to estimate the lateral interaction energy (kink energy, $J$) at the step edges (25). The relationship $\tilde{\beta} \approx \frac{k_B T}{2 a_\perp} \exp(J/k_B T)(1 - \exp(-J/k_B T))^2$ yields estimates $J_{CE} \sim 32$ meV (43 meV), and $J_S \sim 43$ meV (56 meV). The estimated magnitude of $J_S$ is much smaller than the expected value of a hydrogen bond, ~0.3-0.4 eV. In addition, the ratio of $J_S/J_{CE}$ is also much smaller than the expected value of 4-6 for the ratio of hydrogen bonds to quadrupolar interactions. Thus the measured values of the step stiffness are clearly inconsistent with the anisotropy that would be expected for a model based solely on intermolecular interactions between ACA molecules.

Also, given the values for the edge mobilities, $\Gamma_a$, the average time between molecular attachments/detachment events $\tau_a$ can be expressed as (28):

$$\Gamma_a = \frac{a_{//}^2 a_\perp}{\tau_a}, \qquad (7)$$

where $a_{//}$ is the molecular dimension along the boundary direction and $a_\perp$ the molecular dimension perpendicular to the boundary direction. This yields values of $\tau_a \approx 34$ ms (17 ms) for CE boundaries and 18 ms (9 ms) for S boundaries. Using a simple estimate of the attachment rate as $1/\tau_a = \nu\exp(-E_a/k_BT)$, with the frequency factor $\nu$ set to a nominal value of $10^{13}$/s, yields estimates of the activation energies of 0.67 and 0.68 eV, suggesting that equally strong ruptures are needed to reconfigure the edges of the two types of boundaries.

**Theoretical Model of Molecular Interactions**

The asymmetry of the chain bonding in the ordered phase, shown in Fig. 1, should impose a substantial difference in the behavior of the S and CE edge boundaries. Thus the similarity of the measured thermodynamic and kinetic values, specifically the step stiffness and mobility listed in Table I, for these two boundaries is initially quite surprising, and has very interesting physical consequences (4, 36, 37). Here we will derive a specific model for the interactions between ACA molecules, modeled as pairwise interactions within a lattice gas framework, and test the model's ability to reproduce the experimentally measured edge free energies (stiffnesses).

In the simple lattice model discussed in the Introduction, each site on a square lattice is either vacant or occupied by a single ACA molecule. Chain-like structures arise from the anisotropic nearest neighbor interactions in which adjacent molecules in one direction (see Fig. 1) have a large favorable energy $-\varepsilon_{CE}$ due to the H-bonding, while neighbors in the opposite (side) direction have a higher energy $-\varepsilon_S$, with $\varepsilon_{CE} \gg \varepsilon_S$ due to the weaker quadrupolar interaction. ACA-substrate interactions are the same for all molecules in this model, which maps onto the usual spin ½ Ising model with anisotropic ferromagnetic coupling between nearest neighbor spins (24, 25). As shown above, the experimental observations simply do not follow the intuitive expectations of highly asymmetric island shape and edge stiffnesses that follow from this model. A final, and important, problem with this simple lattice gas model is that it predicts a low density in the disordered phase except very near the critical temperature, in contrast with the large density (~ 0.3 ML) observed experimentally (15, 16).

We resolve the apparent discrepancies, while still staying within a lattice gas framework, by taking into account the molecular tilt orientation. As noted in Fig. 1, the strong H-bonds along the chain require alternating tilt-orientations of adjacent ACA molecules. As illustrated in Fig. 5, we can treat these distinct orientations as different species of a multi-component lattice gas, where sites are now either vacant or singly occupied by tilted molecular species U ("up" shown as red/dark gray in Fig. 5) or D ("down" shown as blue/medium gray). This leads to a three-component (U, D, and vacancy) lattice gas that maps onto an anisotropic version of the Blume, Emery, Griffiths (BEG) or spin 1 Ising model (36, 37). There have been many studies (36) of different versions of the BEG model in other contexts, though not with parameters expressing the competition between anisotropic bonding (intermolecular) and substrate (single particle field) terms needed here.

As per the physical model of Fig. 1, we assume that strong H-bonds with energy – $\varepsilon^c_{UD}$ form only if a U is next to a D in the chain direction. In the side direction, the quadrupolar interaction $-\varepsilon^s_{UD}$ is much less favorable, as are energies $-\varepsilon_{UU}$ and $-\varepsilon_{DD}$ of (non-H-bonded) UU or DD pairs in the chain direction, which we take equal to those in the side direction for simplicity (i.e., we have set $\varepsilon^s_{UU} = \varepsilon^c_{UU} = \varepsilon_{UU}$ and $\varepsilon^s_{DD} = \varepsilon^c_{DD} = \varepsilon_{DD}$). Experimentally (see Fig. 1b), the lateral ordering of the chains provides evidence that the symmetrical side interactions ($-\varepsilon_{UU}$ and $-\varepsilon_{DD}$) are somewhat more favorable than the unsymmetrical $-\varepsilon^s_{UD}$.

It seems quite plausible that different conformers (tilt-orientations (38-41)) will have different effective interaction energies with the substrate. We arbitrarily designate U as the tilt-orientation with a more favorable interaction $-\varepsilon_U$ with the substrate. Then the difference in total energy between a H-bonded UD pair and a weakly bonded UU pair in the chain direction results from a competition between a favorable intermolecular energy $\varepsilon_{UU} - \varepsilon^c_{UD}$ and a less favorable substrate energy $\varepsilon_U - \varepsilon_D$. To allow independent control of the density in the disordered phase, we introduce another component G to describe the dominant orientation in the disordered phase, shown as green (light gray) in Fig. 6. An isolated G molecule is supposed to have an optimal orientation with respect to the substrate, thus an even more favorable substrate interaction energy $-\varepsilon_G$ than the $-\varepsilon_U$ of the favorable U orientation. The substrate interaction energy, $\varepsilon_G$, physically could include some relaxation energy of the substrate around an isolated G molecule, some of which could be lost for two adjacent G molecules (36, 42). This can be represented within a lattice gas framework by a weak repulsive effective intermolecular interaction between a GG pair, $\varepsilon_{GG} < 0$. This contrasts with the weak attractions between UU and DD pairs needed for lateral ordering of the chains.

The overall Hamiltonian for this 4-component lattice gas in its most general form is:

$$-H = \sum_{i=U,D,G} N_i \varepsilon_i + \sum_{i,j} \sum_{\alpha=s,c} N_{ij}^\alpha \varepsilon_{ij}^\alpha . \tag{7}$$

Here $N_i$ is number of molecules of species $i$ (= U,D,G) and $\varepsilon_i$ the corresponding substrate energy, $N_{ij}^\alpha$ is the number of nearest neighbor pairs of species $i$ and $j$ in the chain ($\alpha = c$) or side ($\alpha = s$) directions, and $\varepsilon_{ij}^\alpha$ the associated pair interaction energies. The only attractive interaction for UD pairs occurs in the chain direction, $\varepsilon_{UD}^C$, corresponding to the strong H-bonds.

There are many parameters to be determined, as illustrated in Fig. 5, but we can use the experimental observations to help choose physically relevant values. An important relationship is obtained by requiring that the T = 0 boundary energies per unit length along the straight S and CE boundaries be essentially the same. Using Eq. (7), it is easy to calculate the total energy at T = 0 of a large rectangular island containing a fixed number of molecules $N_{TOT} = N_S \times N_{CE}$, made up of $N_{CE}$ perfect laterally ordered chains of length $N_S$ molecules:

$$\begin{aligned}-E_{TOT} &= (N_S - 1)N_{CE}\varepsilon_{UD}^c + \frac{1}{2}(N_S + 1)(N_{CE} - 1)\varepsilon_{UU} + \frac{1}{2}(N_S - 1)(N_{CE} - 1)\varepsilon_{DD} \\ &+ \frac{1}{2}(N_S + 1)N_{CE}\varepsilon_U + \frac{1}{2}(N_S - 1)N_{CE}\varepsilon_D\end{aligned} \tag{8}$$

Both CE boundaries will optimally contain U rather than D molecules because of the more favorable substrate interaction, so $N_S$ is odd. Minimizing the energy with respect to $N_S$ with $N_{TOT}$ held constant, and assuming for simplicity $\varepsilon_{UU} = \varepsilon_{DD}$, gives the relation

$$\frac{N_S}{N_{CE}} = \frac{\varepsilon_{UD}^c - (\varepsilon_U - \varepsilon_D)/2}{\varepsilon_{UU}}. \qquad (9)$$

Thus in this multicomponent model, the anisotropy in the T = 0 island shape arising from the strong H-bond interaction $\varepsilon_{UD}^c$ relative to $\varepsilon_{UU}$ is effectively reduced by the difference in the substrate interactions $(\varepsilon_U - \varepsilon_D)/2$. To agree with the experimental observation, the minimum energy should occur for a square island with length $N_S a_S = N_{CE} a_{CE}$, where $a_S / a_{CE} = 1/\sqrt{3}$, and thus $N_S/N_{CE} = \sqrt{3}$. For the case here, with the bonding asymmetry set by the ratio of hydrogen-bond to quadrupolar interaction strength, $\varepsilon_{DD} = \varepsilon_{UU} \approx \varepsilon_{UD}^C / 5$, Eq. (9) yields the estimate $\varepsilon_U - \varepsilon_D \approx 1.3 \varepsilon_{UD}^c$. In other words, to produce a symmetric domain shape consistent with the experimental observation despite the very anisotropic pair interactions requires a substrate binding energy difference comparable to the strongest intermolecular interaction.

Another constraint from experiment arises from the equilibrium between the disordered gas phase, with a rather high density of isolated G molecules, and the chain phase. At low temperature, addition or removal of an H-bonded UD pair at either the S or CE boundary of a chain island creates an effective "kink" (repeatable excitation unit). Any significant channel for interface fluctuations must allow exchange between H-bonded UD pairs at such a kink and two distinct G molecules. Efficient exchange demands a small difference in the total energies of the two configurations, so that

$$2\varepsilon_G \approx 2\varepsilon_{UD}^c + \varepsilon_{UU} + \varepsilon_{DD} + \varepsilon_U + \varepsilon_D. \qquad (10)$$

Using the estimates above in Eq. (10) yields $\varepsilon_G - \varepsilon_U \approx .55\varepsilon_{UD}^c$. The room temperature desorption of submonolayer ACA films from Ag(111) on a 10-hour timescale supports a value of $\varepsilon_G$ value around 0.65 eV (16).

Using these estimates as a starting point, we performed Monte Carlo simulations of the lattice gas model to determine parameter values consistent with the experimental observations. The simulations used a grand ensemble where changes in the total number of molecules of any species $i = $ (G, U, D) on the substrate are controlled by a common chemical potential and the change in local energy on addition or removal of the particle as determined by the Metropolis criterion. Alternatively, a molecule of one species can directly convert to another molecular species with a probability based only on the local change in energy. By testing a range of relative parameter values, we arrived at the values given in Table 2. The substrate binding energy differences follow the trends expected from the T = 0 estimates discussed above, with $\varepsilon_U - \varepsilon_D = 1.29\varepsilon_{UD}^c$ and $\varepsilon_G - \varepsilon_U = .46\varepsilon_{UD}^c$.

Given the values of $\varepsilon_{UD}^C$, this corresponds to orientationally induced changes in the substrate interaction energy from the strongest value of 0.65 eV to only 2.5 meV. This model suggests that half the ACA molecules (the unfavorable D molecules) in the ordered structure are primarily held in place by intermolecular interactions. The chemisorption of aromatic molecules such as ACA is dominated by pi-bonding to the substrate. However, nitrogen-containing heterocycles are known to adopt tilted-configurations via additional nitrogen lone-pair interactions with the substrate (38-41). Thus a reasonable hypothesis is to assign the most favorable G state to a planar configuration, the U state to the configuration with the N atom down, and the least favorable D state to the configuration with the carboxyl group down.

Fig. 6 shows equilibrated Monte Carlo configurations of the lattice gas model with parameters from Table 2 for a state with fixed temperature of 371 K, where finite strips of the solid, mainly the UD chain phase, are in equilibrium with a dense vapor of mostly isolated G molecules. The similarity in fluctuations at the orthogonal boundaries is evident, with rms widths $w_S$ = 2.23 nm and $w_{CE}$ = 2.20 nm, consistent with the experimental ratio close to 1. We use the Monte Carlo configurations to directly determine the step stiffness $\tilde{\beta}$ at each boundary from the equilibrium spatial correlation function $G(y) = \langle (x(y+y_o) - x(y_o))^2 \rangle$ (43, 44). According to the capillary wave model, in a large system of size $L$ with periodic boundary conditions this should equal (45)

$$\tilde{G}(y) = \frac{k_B T}{\tilde{\beta}} y(1 - y/L) \qquad (11)$$

for $0 \le y \le L$. Figure 7 shows the position correlation function for both the side and end boundaries with the fit to Eq. (11). This yields accurate absolute estimates for the stiffnesses, $\tilde{\beta}_{CE}$ = 103 meV/nm and $\tilde{\beta}_S$ = 102 meV/nm for this parameter set. Their ratio $\tilde{\beta}_s/\tilde{\beta}_{CE} = 1.01$ is in reasonable agreement with the experimental ratio, $\tilde{\beta}_s/\tilde{\beta}_{CE} = 1.2 \pm 0.3$. The absolute value of the MC stiffness is somewhat larger than the experimental upper limit (~ 80-90 meV/nm), suggesting that the true H-bond strength may be somewhat smaller than the value of 0.37 eV assumed in the simulation.

**Discussion and Conclusions**

Boundaries within ACA thin films fluctuate dynamically with rms displacements of 2.0 nm and underlying molecular-exchange time constants of 10-30 ms at room temperature (see Table I). The fluctuations are well described by the equilibrium properties of line boundaries with non-conserved thermal excitations balanced by a restoring line tension. We expect that such fluctuations will influence electron transport properties and impact electronic applications of organic thin films. For instance reduced carrier transport across a domain boundary, or the formation or recombination of electron-hole pairs at a boundary will generate a time-dependent signature due to the fluctuations in length of the boundaries, which evolve with characteristic time constants $\tau_k \approx \frac{k_B T}{\tilde{\beta} k^2 \Gamma_a}$, where $k$ is the wavenumber corresponding to the mode of fluctuation (46). For the ACA thin films, the signature of this time dependence will be a noise component with frequency dependence $f^{-1/2}$ (46, 47), with the largest frequency limited by the underlying molecular exchange times $\tau_a$. For the edge boundaries of ACA, the upper limit of the frequency contribution will thus be in the range of 30-100Hz. Such noise signatures may have interesting consequences, such as stochastic resonance, in designing future novel nanoelectronic device properties (48-52). It is possible to estimate the value of the characteristic time constant $\tau_a$, using the standard rate equation $1/\tau_a = \nu \exp(-E_a/kT)$ if the activation energy $E_a$ and frequency factor $\nu$ for motion are known or can be approximated. Relevant measured and calculated values for substrate mobility based on step motion have been measured (28, 29, 31, 44, 53), but there are relatively few studies of boundary motion for molecular species (54, 55). However, activation energies can be estimated based on molecular diffusion and molecular interactions(56-59). The boundaries in the ACA films also displayed surprising structural characteristics — they are isotropic in island shape, line tension and fluctuation time constant despite the strongly anisotropic intermolecular interactions, with a ratio of orthogonal H-bond to quadrupolar interactions $\varepsilon_{UD}^C/\varepsilon_{UU} \approx 5$. We address this by adapting the BEG model to incorporate an additional energetic effect for the ACA molecule, where configurations with alternate tilt angles are observed experimentally in the condensed phase. The energy costs of the tilted configurations are introduced as single-particle field effects, and are modeled as discrete, but interchangeable, molecular components, parameterized as $\varepsilon_U$ and $\varepsilon_D$. Physical analysis of the low-temperature limit of the model shows that the effective anisotropy of the overlayer can be changed dramatically, as represented by the shape ratio: $\frac{N_S}{N_{CE}} = \frac{\varepsilon_{UD}^c - (\varepsilon_U - \varepsilon_D)/2}{\varepsilon_{UU}}$. Monte Carlo simulation using the model yields quantitative agreement with the experimental observations, and suggests that molecular tilt can reduce the molecular-substrate interaction strength from its optimum value near 0.65 eV to near zero.

The effect of molecular tilt is frequently observed in N-heteroaromatics (38-41), and thus the symmetry modifications observed for ACA should also be common. By considering the shape ratio shown above, it is easy to predict the consequences of modifying the intermolecular interactions. In the ACA case, where the asymmetry of the molecular interactions $\varepsilon_{UD}^C/\varepsilon_{UU}$ is large, the tilt effect effectively symmetrized the boundaries, thus creating compact domains. Conversely, modifying the molecular structure to reduce the interaction asymmetry (e.g. reduce $\varepsilon_{UD}^C/\varepsilon_{UU}$) by increasing the molecular interactions orthogonal to the hydrogen-bond direction would exaggerate the asymmetric shape ratio for the same relative tilt-effect. Beyond this one example, many other combinations can be evaluated using this model, effectively yielding a molecular basis for predicting thin film morphology. The ability to tune domain shape will allow the systematic design of organic thin film systems that will self-assemble boundary configurations favorable for charge transport and recombination (1, 60).

**Materials and Methods**

The substrates, Ag thin films on mica, are made by thermal deposition, as described previously (32), and transferred into a UHV chamber (base pressure ~ 3 × $10^{-11}$ torr). After several sputtering and annealing cycles, atomically clean Ag(111) surfaces are obtained, as confirmed by both LEED and STM. By carefully controlling annealing conditions, large defect-free terraces (> 1 µm) form on the Ag films.

Methods for the preparation of well-controlled ACA films have been determined previously (14-16). In a preparation chamber (base pressure ~ 5 × $10^{-9}$ torr) contiguous to the main UHV chamber, ACA in powder form is placed in a quartz effusion cell. After fully degassing the cell at ~ 380 K, ACA molecules are thermally deposited at 393 K onto the Ag substrates, which are held at room temperature. The deposition rate is ~ 0.3 ML/min, calibrated by a quartz microbalance and subsequent STM measurement. When the ACA coverage is larger than 0.3 ML, ACA molecules coexist in ordered and disordered phases.

STM imaging was performed using care to avoid tip-induced effects as demonstrated in earlier work (32, 61, 62). The tunneling current is ~ 40 pA at sample bias ~ 0.90 V. These mild tunneling conditions were carefully assessed to assure that the tip-sample interactions do not affect the measurements. To observe boundary fluctuations, we use repeated STM scans across a fixed position at the boundary, along the direction approximately perpendicular to the boundary direction (55, 63). The time interval between sequential scans is 51.2 ms and the total measurement time is 102.4 s with 2000 lines for each temporal pseudoimage. Images are recorded in both the forward and backward scanning directions, and analyzed separately. Typically 15 such forward/backward data sets are measured for each boundary investigated. To extract the position of the boundary as a function of time $x(t)$, we flatten the temporal pseudo-images by tilting the images to one level of the two phase regions, and then identify the boundary at which the height is midway between the two phases.

**Acknowledgments**

This work has been supported by the UMD NSF-MRSEC and its shared facilities under grant # DMR 05-20471. Infrastructure support from the UMD Center for Nanoscience and Advanced Materials (CNAM) and the UMD NanoCenter is also gratefully acknowledged.

**Table 1**: Synopsis of values ($1/n$, $w$, $G_0$ and $\tau_c$) directly determined from the distribution and correlation functions, and the derived values ($\tilde{\beta}$, $\Gamma_a$, $\tau_a$) obtained using the limiting values of $L = 60$ nm and $L = 120$ nm. All values correspond to T = 300 K.

|  | CE-boundary (parallel to $[11\bar{2}]$) | S-boundary (parallel to $[1\bar{1}0]$) |
|---|---|---|
| $1/n$ | $0.516 \pm 0.087$ | $0.512 \pm 0.046$ |
| $w$ (nm) | $2.04 \pm 0.04$ | $2.01 \pm 0.08$ |
| $G_0$ (nm$^2$s$^{-1/2}$) | $3.76 \pm 0.25$ | $3.70 \pm 0.72$ |
| $\tau_c$ (s) | $3.59 \pm 0.30$ | $2.73 \pm 0.43$ |
| $L$ (limits) (nm) | 60 – 120 | 60 –120 |
| $a_{//}$ (nm) | 1.001 | 0.578 |
| $a_{\perp}$ (nm) | 0.578 | 1.001 |
| $\tilde{\beta}$ (meV/nm) -lower limit | $39.1 \pm 4.2$ | $45.5 \pm 12.4$ |
| upper limit | $78.2 \pm 8.4$ | $91.0 \pm 24.8$ |
| $\Gamma_a$ (nm$^3$/s) - lower limit | $16.8 \pm 0.4$ | $19.0 \pm 2.2$ |
| upper limit | $33.6 \pm 0.8$ | $38.0 \pm 4.4$ |
| $\tau_a$ (s) - upper limit | $0.034 \pm 0.001$ | $0.018 \pm 0.002$ |
| lower limit | $0.017 \pm 0.0004$ | $0.009 \pm 0.001$ |

**Table 2**: Parameters determined from MC simulations. A reasonable value of 0.37 eV was chosen for the H-bond energy $\varepsilon_{UD}^c$ and the ratio of $\varepsilon_{UD}^c$ to $\varepsilon_{UU} = \varepsilon_{DD}$ was set to an assumed value of 5. The symmetry balance and step stiffness are mainly determined by this ratio and the difference $\varepsilon_U - \varepsilon_D$ as given in Eq. (10). We constrained our final parameter set to satisfy Eq. (10) for square T=0 islands, as suggested by experiment. The density in the gas phase increases with $\varepsilon_G$, which was chosen to give a high density gas phase at experimental temperatures.

| Value | Parameter |
|---|---|
| 0.65 eV | $\varepsilon_G$ |
| 0.48 eV | $\varepsilon_U$ |
| 0.37 eV | $\varepsilon_{UD}^c$ |
| 0.075 eV | $\varepsilon_{UU}, \varepsilon_{DD}$ |
| 0.0 eV | $\varepsilon_D, \varepsilon_{GU}, \varepsilon_{GD}, \varepsilon_{UD}^s$ |
| -0.02 eV | $\varepsilon_{GG}$ |

**Figures**

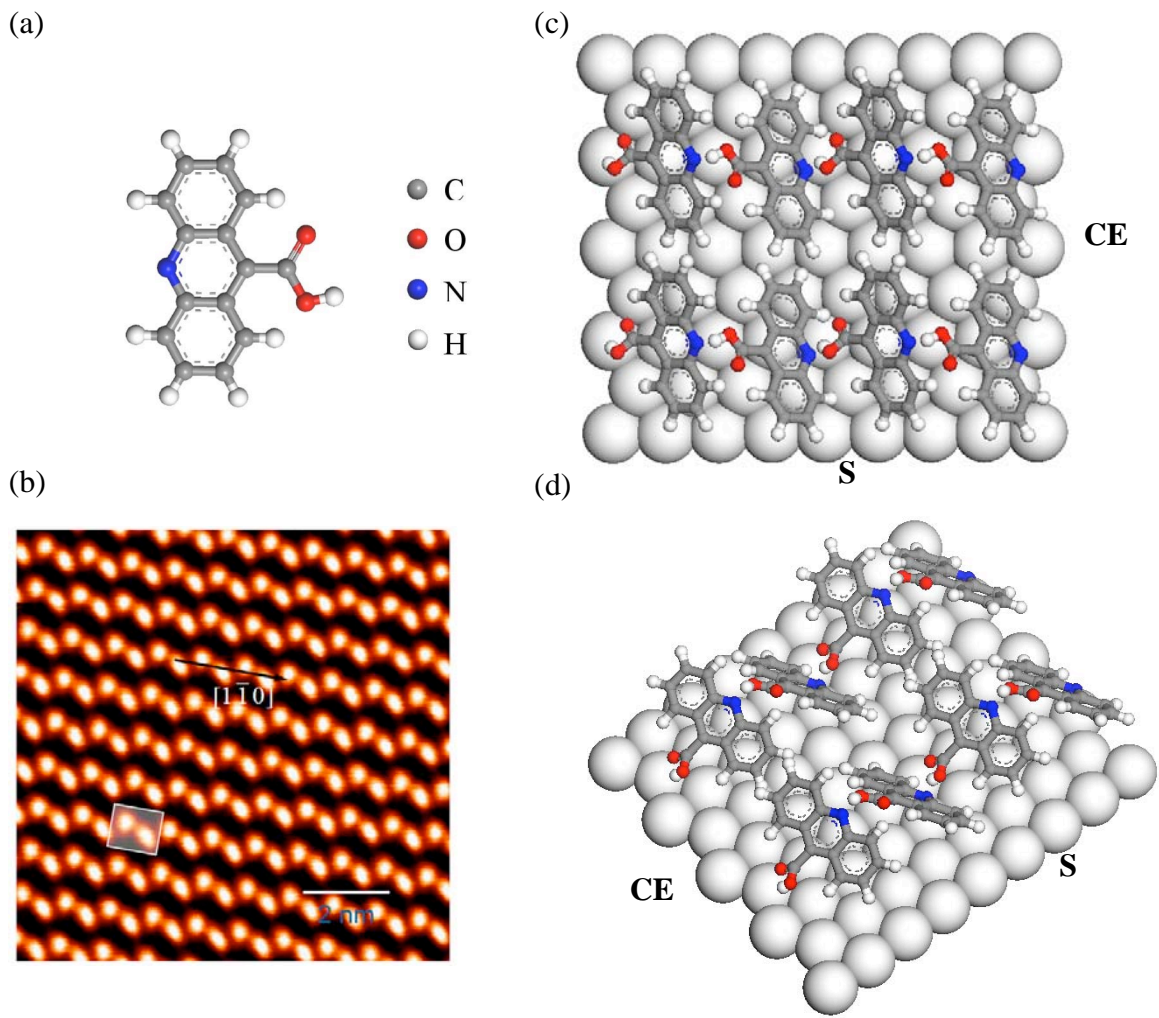

Fig. 1 (Color online) Structural models and images of ACA on Ag(111). (a) Molecular structure of ACA,. Arrangement of ACA molecules in the ordered phase (b) as measured using STM and (c-d) corresponding molecular models determined previously using STM, IRRAS and XPS measurements (15, 16) (top view and perspective view). Alternating ACA molecules along the head-to-tail hydrogen-bonded chains are tilted with respect to the substrate by 45 degrees. The molecular dimensions are $a_S \sim 2a$ in the $[1\bar{1}0]$ direction and $a_{CE} \sim 2\sqrt{3}\,a$ in the $[11\bar{2}]$ direction, where $a = 0.289$ nm is the near-neighbor distance on Ag(111).

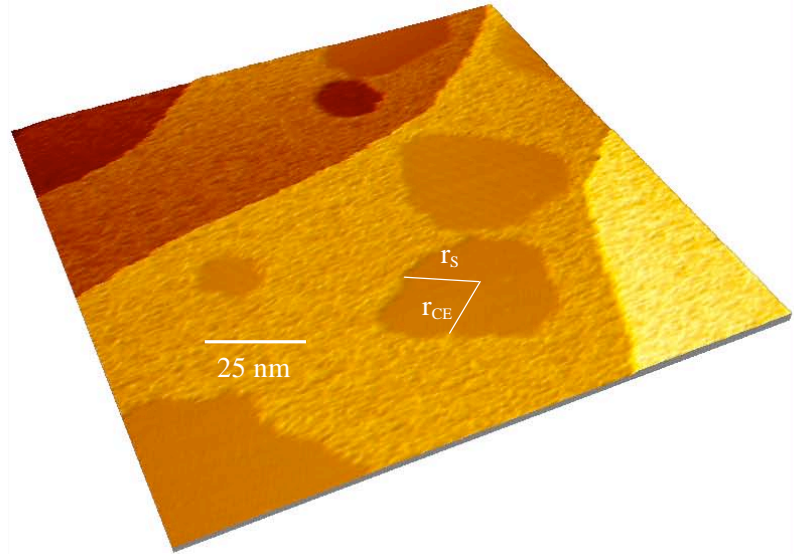

Fig. 2 STM images of ordered islands and surrounding disordered phase. $r_{CE}$ and $r_S$ indicate the distance from center to edges, with aspect ratio $r_{CE}/r_S$ close to 1. The scanning conditions are $V_s = -0.65$ V and I = 34 pA.

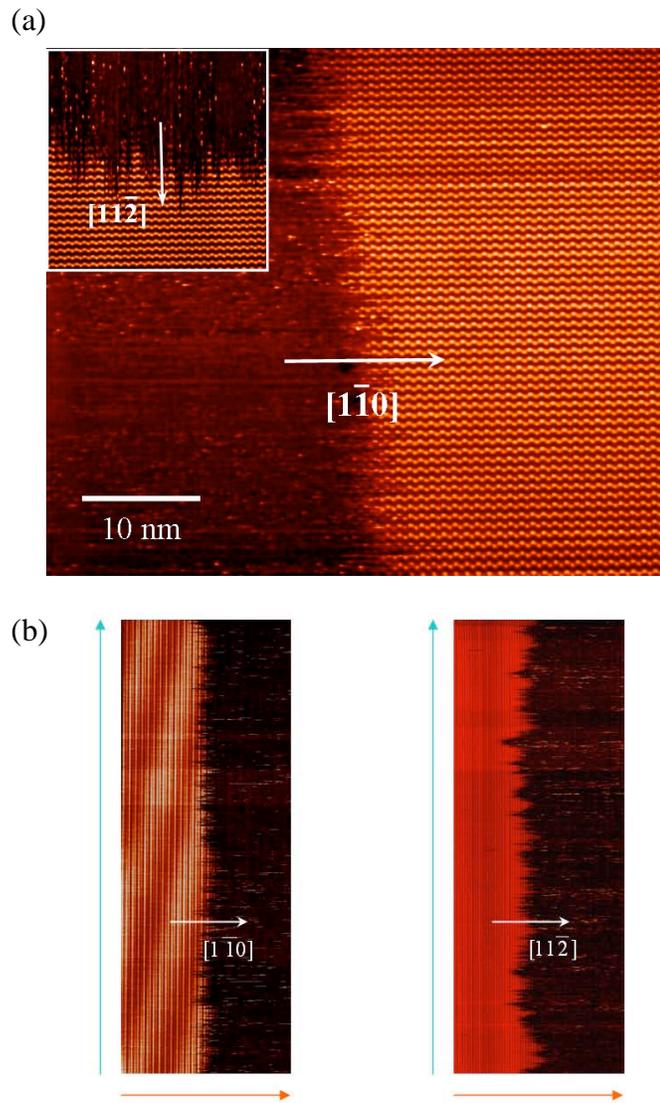

Fig. 3 (color online). Measurements of boundary fluctuations. The white arrows indicate the scan directions. (a) STM images of boundaries of ordered and disordered phases. The boundary is perpendicular to the $[1\bar{1}0]$ direction and the scan direction is along $[1\bar{1}0]$ direction (CE boundary); (inset) boundary perpendicular to $[11\bar{2}]$ direction with the scan direction is along the $[11\bar{2}]$ direction (S boundary). The scanning conditions are $V_s = -0.95$ V and I = 47 pA. (b) Pseudo-images of boundaries fluctuations, with line scan size 50 nm (horizontal axis), line scan time 51.2 ms, and total measurement time 102.4 s (vertical axis) for 2000 lines. (left) boundary perpendicular to the $[1\bar{1}0]$ direction with the scan direction along the $[1\bar{1}0]$ direction (CE boundary); (right) boundary perpendicular to the $[11\bar{2}]$ direction with the scan direction along the $[11\bar{2}]$ direction (S boundary).

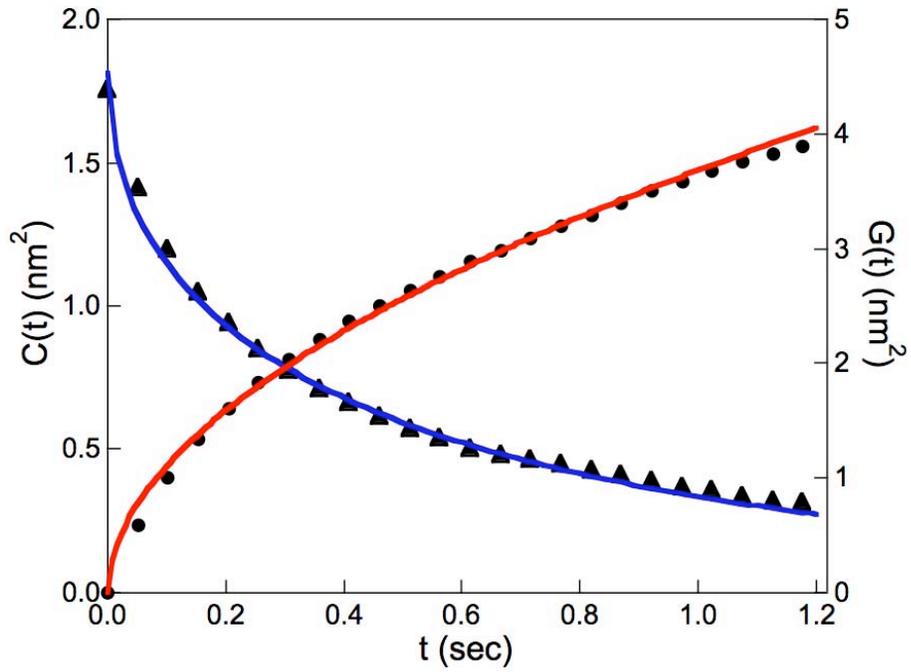

Fig. 4 (Color online). Typical correlation function $G(t)$ (black solid circles, right axis, CE boundary) and autocorrelation function $C(t)$ (black solid triangles, left axis, S boundary) data. The red curve is the fit for $G(t)$ using Eq. (4), and extracts the best fit values, $G_0 = 3.86 \pm 0.28$ nm$^2$, $1/n = 0.52 \pm 0.07$. The blue curve is the fit for $C(t)$ using Eq. (6), and extracts the best fit values, $C(0) = 1.81 \pm 0.36$ nm$^2$ and $\tau_c = 1.77 \pm 0.49$ s.

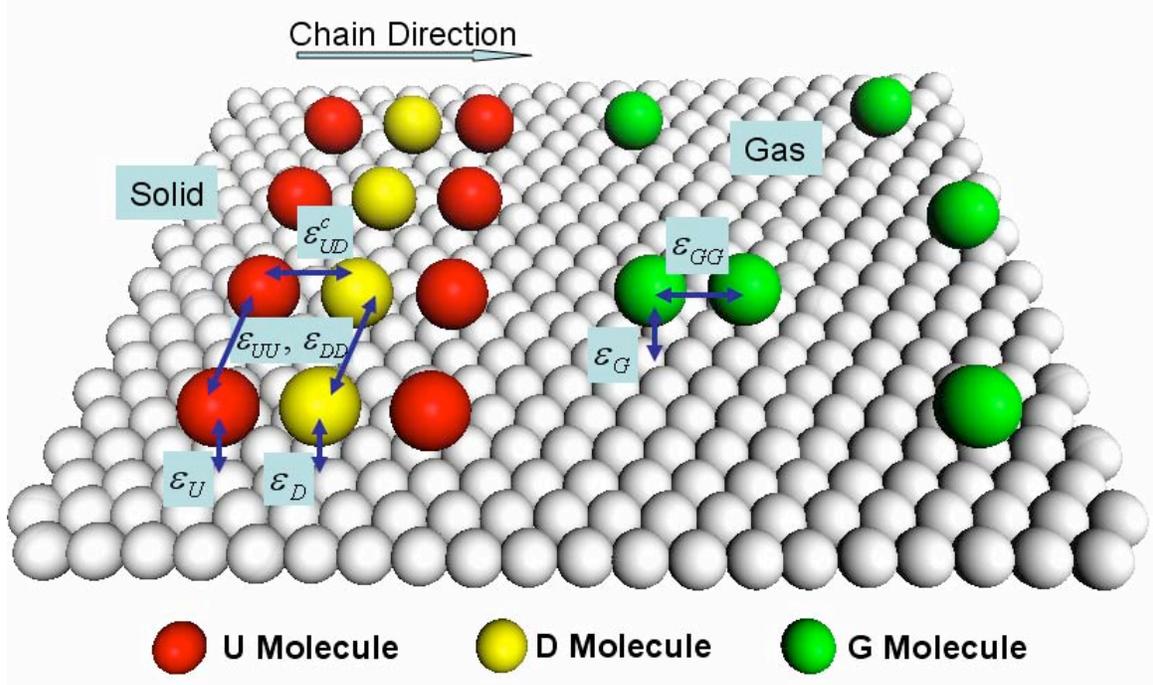

Fig. 5 (Color online). Schematic illustration of molecular system with variable substrate interactions as well as intermolecular interactions. U and D correspond to up/down orientations of the molecular tilt as shown in Fig. 1, while G corresponds to an orientation with most favorable interaction with the substrate, possibly perfectly horizontal. The parameters $\varepsilon_{ii}$ correspond to lateral interactions between molecules of orientation i = U, D or G, and the parameters $\varepsilon_i$ correspond to interaction of molecules of orientation i = U, D or G with the substrate.

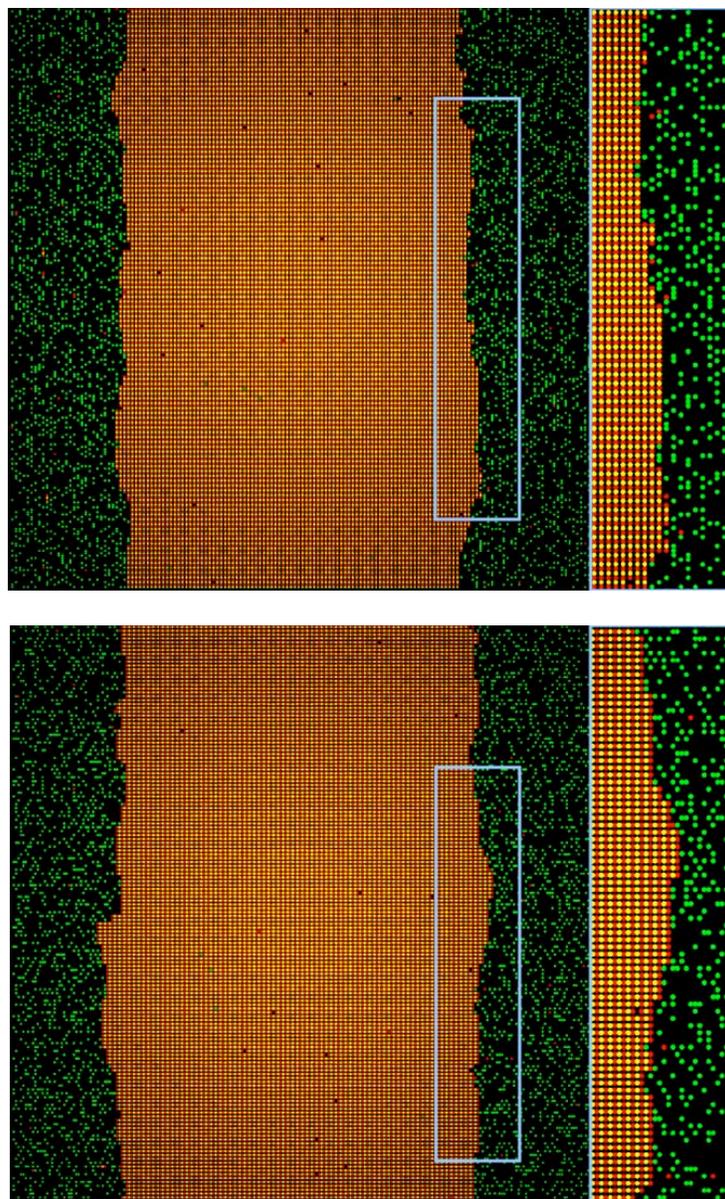

Fig. 6 (Color online). Images of Monte Carlo simulations. A condensed strip with side (S) boundaries in equilibrium with the dense gas phase is shown in the upper panel, and one with end (CE) boundaries is shown in the lower. The areas enclosed in the boxes are shown in expanded view to the right. Both simulations use a (139 nm x 150 nm) (240$a_S$ x 150 $a_{CE}$) system with periodic boundary conditions. Simulations were initiated with two perfectly straight boundaries, and the images shown are after 300,000 sweeps. The simulation temperature 371K was chosen somewhat higher than experiment to permit more rapid equilibration between the phases and the simulations determined the coexistence chemical potential as −0.675 ev.

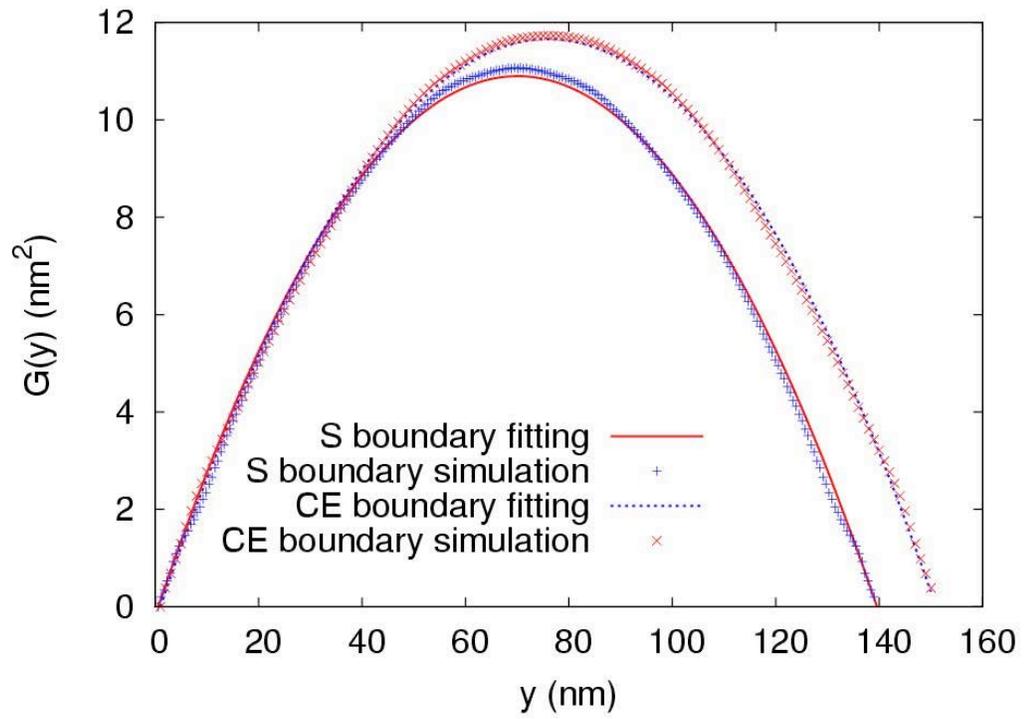

Fig 7 (color online). Simulation of $\tilde{G}(y)$ and fits for $\tilde{\beta}$ using Eq. (12) for the boundaries shown in Fig. (7). For the CE boundary, $L$ = 150 nm, and the best fit value $\tilde{\beta}$ = 103 meV/nm. For S boundary, $L$ = 139 nm, and fit value $\tilde{\beta}$ = 102 meV/nm.